# To the question of gravitational emission's linear spectrum identification, as of an emission of the same level with electromagnetic.

**S. I. Fisenko, I. S. Fisenko**

"Rusthermosinthes" JSC
Moskva-City, "Tower 2000"
T. Shevchenko 23A - "B", Moscow, 121151, RUSSIA
Phone: +7 (495) 255-83-64, Fax: +7 (495) 255-83-65
E-mail: StanislavFisenko@yandex.ru

## Abstract

The notion of gravitational emission as an emission of the same level with electromagnetic emission is based on the proven fact of existence of electron's stationary states in its own gravitational field, characterized by gravitational constant $K=10^{42}G$ (G – Newton's gravitational constant). The transition to stationary states is to cause gravitational emission of an electron that is characterized exactly by constant K. To the development of these results this paper reviews the algorithm of this sort of emission's spectrum identification in the experiments with electron beams.

## 1. Initial formalism

Generally covariant equations form in gravity relativity theory, as it is well-known, is as follows:

$$R_{ik} - \frac{1}{2} g_{ik} R - \Lambda g_{ik} = \chi T_{ik} \qquad (1)$$

In these equations $\chi$ is the constant, connecting the space-and-time geometric property with the distribution of physical material, so the origin of the equations is not connected with numeric limitation of $\chi$ quantity value. Only the necessity to correspond classical Newtonian gravitation theory brings us to numeric value $\Lambda = 0, \chi = 8\pi G/c^4$, where *G is* Newtonian gravitation constant. The equations with the defined constants are the equations of the Einsteinian general relativity theory (GRT). The equations (1) are common mathematical form of gravitational field equations, corresponding to the equivalency principle and general covariance axiom. The equations in form (1) were acquired simultaneously with Einstein but independently from him by Hilbert [1]. In [2, 3, 4] there was made a simple and at the same time strict assumption of the existence of such numeric values of gravitational constant K and constant $\Lambda$ in quantum sphere, that bring to stationary states in own gravitational field, and these are already emitters of gravitational field with Newtonian gravitational constant. The very numeric values of K and $\Lambda$ are estimated independently, exactly within this approach. Herewith we make reference to A. Salam [5], as he was one of the first to pay attention to Newtonian gravitational constant's numeric value does not conform to quantum level. He was the one to propose the concept of "strong" gravity, that was based on the assumption of f-mesons spin 2 existence, that form SU(3)- multiplet (described by Paul-Firz). It was proven that a possibility of a different link constant along with Newtonian one does not contradict the observed effects [5]. Due to a number reasons these approach was not developed further. As it is clear now, this numeric value of "strong" gravity constant is to be used in equations (1) with $\Lambda \neq 0$. Besides, precisely with $\Lambda \neq 0$ stationary solutions of general Einsteinian equations can be found, which was noticed by Einstein himself, but after discovery of functionary solutions with $\Lambda = 0$ by A. Friedman [6], the modern shape of the GRT was finally formed. The decisive argument of GRT to equal $\Lambda$-element to zero is the necessity of right limit passage to Newtonian gravitation theory.

In the simplest (from the point of view of the original mathematical estimations) approach the problem on steady state in own gravitational field (with constants K and $\Lambda$) is solved in [3, 4]. From the solution of this problem it becomes evident:

a) With numeric values of $K \approx 5.1 \times 10^{31}$ Nm²kg$^{-2}$ и $\Lambda = 4.4 \times 10^{29}$ m$^{-2}$ there is a spectrum of electron's stationary states in proper gravitational field (0.511 MeV …0.681 MeV). The main state is detected electron's rest energy 0.511 MeV.

b) These stationary states are the emitters of gravitational field with *G* constant.

c) The transition to stationary states in own gravitational field causes gravitational emission that is characterized by constant K, and with this is the emission of the same level with electromagnetic (electric charge e, gravitational charge $m\sqrt{K}$). In this respect it is meaningless to say that gravitational effects in quantum area are characterized by G constant, this constant belongs only to microscopic level and it cannot be transferred to quantum level (which is evident from negative results of gravitational waves with G constant registration tests, these do not exist).

d) The existence of stationary states in own gravitational field also completely corresponds the special relativity theory. According to SRT, relativistic link between energy and impulse is broken, if we assume that full electron's energy is defined only by Lawrence's electromagnetic energy [7]. If to expand the situation it is as follows [7]. Energy and impulse of the moving electron (with the assumption that the distribution of the electric charge is spherically symmetrical) is defined by expression:

$$\overline{P} = \overline{V} \frac{\frac{4}{3} E_0 / c^2}{\sqrt{1-\beta^2}} \quad (2)$$

$$E = \frac{E_0 (1 + \frac{1}{3} V^2 / c^2)}{\sqrt{1-\beta^2}} \quad (3)$$

If these expressions were at the same time defining full impulse and full energy, the following relator would take place:

$$E = \int (\overline{V} \frac{d\overline{P}}{dt}) dt \quad (4)$$

However this relator cannot take place as integral in the right part equals to:

$$\frac{\frac{4}{3} E_0}{\sqrt{1-\beta^2}} + const \quad (5)$$

If we find that impulse contrary to the energy has strictly electromagnetic character, then to $E'$ of the moving and full energy $E'_0$ of the resting electron, and also to rest mass $E_0$, following relators will take place:

$$E' = \frac{E'_0}{\sqrt{1-\beta^2}}, \quad E'_0 = \frac{4}{3} E_0, \quad m_0 = \frac{E'_0}{c^2} = \frac{4}{3} \frac{E_0}{c^2}, \quad (6)$$

Where rest mass $m_0$ is defined by following expression:

$$\overline{P} = \frac{m_0 \overline{V}}{\sqrt{1-\beta^2}} \quad (7)$$

Then from (6) it follows that full energy of resting electron equals to $\frac{4}{3}$ of its Lawrence's electromagnetic energy. Numeric data of the electron's stationary states spectrum in own gravitational field fully correspond to it.

e) The spectrum of electron's stationary states in own gravitational field and transitions to stationary states are represented on Fig. 1. We should notice straight away that numeric value is approximate. The largest inaccuracy belongs to numeric value of the first stationary state $E_1$, but it is more and more accurate coming closer to $E_\infty = 171 keV$.

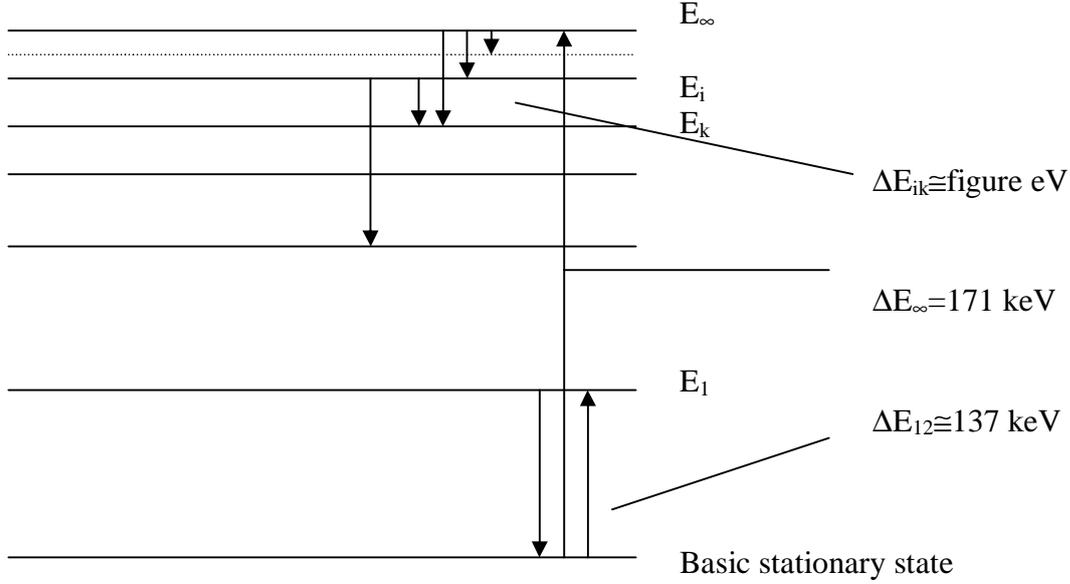

Figure 1. Transition over stationary states of electron in own gravitational.

**2. Macroscopic development of emitted gravitational field's properties.**

For the energies of transition over stationary states in own gravitational field that were mentioned above and energy level width, the only object within which gravitational emission can be actualized as a mass phenomenon is dense high temperature plasma [3]. In DT or DD plasma the conditions of gravitational emission excitation, with its further emission over the whole spectrum from eV to keV of emitted gravitons' energy values, are met. However, along with interim cascade transitions from higher to lower levels (picture 1) there are straight transitions with keV energy levels. Emission of this kind of transitions leaves plasma freely in distinction from eV emission, locked in plasma and then suppressing plasma. It is understood that in such conditions there is no increase in gravitational emission. The best variant would be the appearance of local in time and space high-temperature formations (micro pinch), or anomalously high short time heating of ionic component [8]. Adding multi charge ions to hydrogen plasma (with ions having electron shell's energy levels close to energy levels of electrons in proper gravitational field) will result in keV-level excitation state quenching. Together with cascade transitions this will cause gravitational emission energy transfer to long eV range energy spectrum. With this the presence of impurity of multi electronic atoms in hydrogen plasma will provide locking and increasing gravitational emission in plasma. It should be noted that is well known in optical range (fluorescence) and broadly used for quenching optical range emission lines. This is exactly the mechanism of keeping plasma which is emitted by the gravitational field in addition to the magnetic way of keeping plasma. It can be most clearly seen from the movement of the particle equation in classical approach (and long wave part of the emitted gravitational field complies with it), which is as follows:

$$\frac{d^2 x^i}{ds^2} + \Gamma^i_{kl} \frac{dx^k}{ds} \frac{dx^l}{ds} = 0 \qquad (8)$$

Where $\frac{d^2 x^i}{ds^2}$ is 4-acceleration of the particles and quantity, $m\Gamma^i_{kl} \frac{dx^k}{ds} \frac{dx^l}{ds}$ is «4 –force»

At the same time both actions should obviously take turns in a way which is described in [3, 4]: heating of plasma with the help of magnetic field compression, it's following injection out of the magnetic field for the last phase of compression and keeping plasma emitted by the gravitational field. It is well known that ions of carbon, krypton and especially xenon (especially the latter) are highly emitting systems what results in the loss of plasma's energy while compressing and heating. This is so indeed can be considered to be a drawback of the first phase of plasma compressing and heating. But at the same time this will be an advantage at the final phase, as keeping plasma emitted by the gravitational field and the flowing reaction of synthesis, energy withdrawal from plasma can only take place as a type of electromagnetic energy emission (where it's main part will be shifted to optical zone exactly when flowing of synthesis takes place). We deliberately do not mention the role of fluctuations of plasma's characteristic, coming to nothing more than Maxwellian distribution (to estimate gravitational emission capacity [3, 4]), as a minimum realizable choice. In fact the action itself taking into consideration the fluctuations for example electrostatic inner field strength [9] seems to be more optimistic.

Thus gravitational emission can be exited in dense high temperature plasma and can be magnified in some conditions, but its magnification leads to compression of emitting system. Consequently, when gravitational emission is magnified it is possible to observe not the gravitational emission itself, but only the result of its activity. This circumstance complicates defying of electron's stationary state in own gravitational field spectrum numeric characteristics in experiments with plasma. There are, however, processes that are not connected with magnifying the emission, within which the effects of gravitational emission can be observed per se.

### 3. Identification of electron's gravitational emission spectrum lines in experiments with electron beams.

#### 3.1. Electron beams: gravitational emission spectrum identification.

In technical terms the most accessible is the usage of electron beams, but taking into consideration the condition mentioned in [4] (the smallness of energy level range of beam electrons stationary states in own gravitational field). In a rough approximation the evaluation of stationary states' in own gravitational field energies numeric values gives the following meanings [3, 4]:

$E_1 = 0{,}511 МэВ, E_2 = 0{,}638 МэВ, ..... E_\infty = 0{,}681 МэВ.$

This evaluation comes from a very approximate dependence:

$$E_n = E_0(1 + \alpha e^{-\beta n})^{-1} \qquad n = 1, 2 ..... \infty \qquad (9)$$

где $E_0 = E_\infty = 0{,}681 МэВ, \quad \alpha = 1{,}65; \beta = 1{,}60.$

Undoubtedly, numeric values of the general stationary state energy $E_1 = 0{,}511 МэВ$ are correct, as well as the range of stationary states energies equal 171 keV, i.e. approaching closer to $E_\infty$, energy spectrum numeric values can be estimated more accurately. The largest inaccuracy takes place during the estimation of the first stationary state's numeric value (and of course of a few next ones) and it gets smaller when approaching closer to $E_\infty$. The idea of the experiment is that after stopping electrons beam on the target (together with well-known and thoroughly studied continuous spectrum of electromagnetic bremsstrahlung) the gravitational braking emission with linear spectrum will also take place. Due to recoil during emission, linear spectrum of gravitational emission energy will drift against transition energies $(E_2 - E_1), (E_3 - E_2)$, etc. The value of this drift $\Delta E$ for electrons can be defined very accurately by the following formula:

$$\Delta E \cong 0{,}98 E^2 [МэВ] \qquad (10)$$

From (10) it is clear that energies of gravitational emissions spectrum lines have drifted against transition energies within a frame of 6 keV to 30 keV (and again with great accuracy for the upper energy levels of electron in own gravitational field). Consequently, gravitational emission's linear spectrum is surely in the range of energies (70-140) keV (probably, it is even drifted downwards in energy value). Unlike well-known ways of electromagnetic emission registration (along all of its range), for gravitational emission there are no such registration ways, however gravitational emission's linear spectrum allows us use this feature in comparison with continuous spectrum of Bremsstrahlung in the ranges with concurrent energies. Herewith, overlapping of the characteristic electromagnetic emission lines and Bremsstrahlung continuous range should naturally be taken into consideration. Quantitative character of the experiment is well illustrated by the example of possible emission spectrum of concrete metal anode. The right part of Fig. 2 shows the spectrum of Bremsstrahlung rhodium anode with overlapping characteristic anode emission lines. These data is widely known [10] and are only given for the purpose of comparison. Possible Bremsstrahlung spectrum with overlapping gravitational emission lines is given on the left.

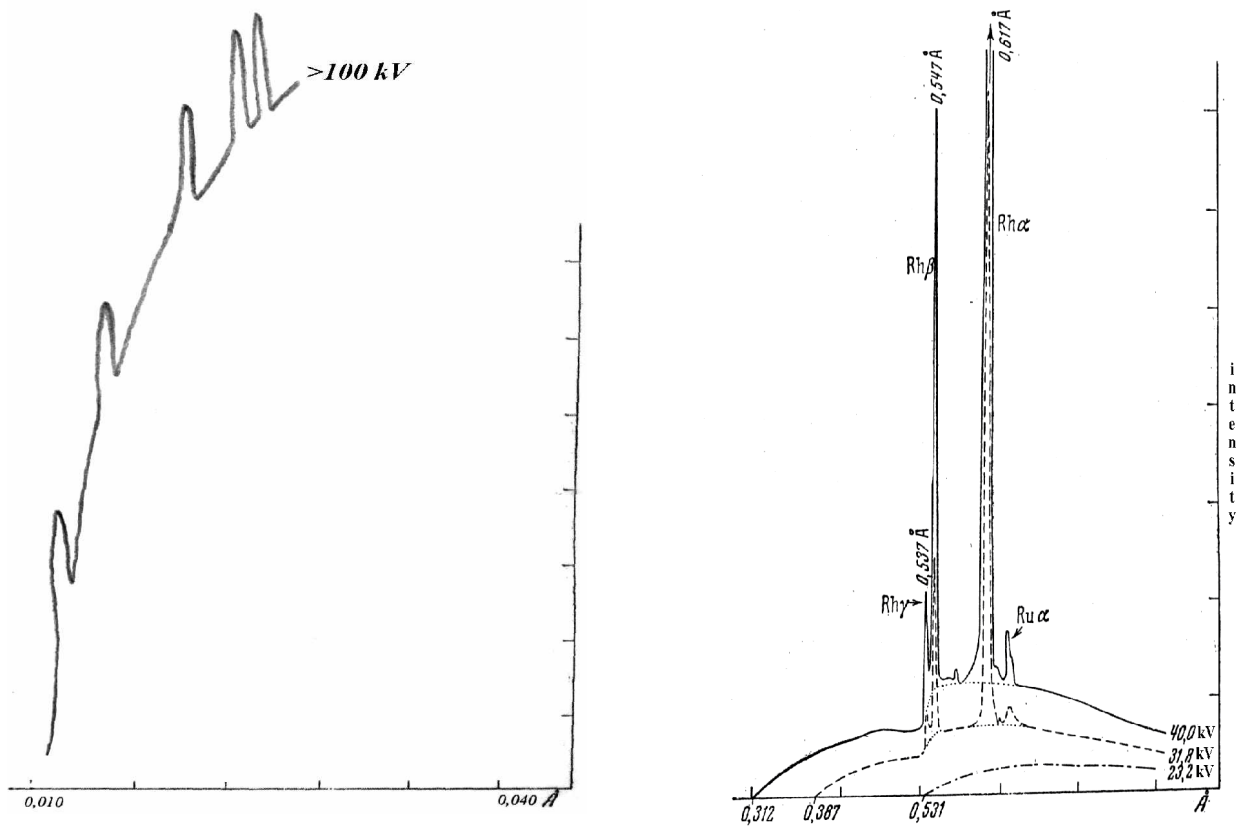

Figure 2. Possible character of gravitational emission lines on rhodium anode (Rh) overlapping with continuous spectrum of Bremsstrahlung (on the left side); on the right side characteristic electromagnetic emission lines also overlapping with continuous spectrum of Bremsstrahlung are given for the purpose of comparison (line α-Ru belongs to ruthenium, which is an impurity element in rhodium sample).

Expected result of gravitational emission spectrum lines registration is based on the notion that gravity quantum influence on detector would be energetically similar to the influence of photon. The main problem would be the defining power of the detector, taking into consideration small width of gravitational emission's lines. Practically, with this will be connected the main difficulty with registration of lines within long wave range (0.01 – 0.04) Å, corresponding to electron's gravitational emission spectrum. It is also understood that calculations should be carried out at different values of accelerating electron's potential, but in any case providing emission range (0,01 – 0,04)Å. And as the character of used metal anode's characteristic emission spectrum is well-known, it will allow evading mistakes during identification of registered lines, as lines of gravitational emission's spectrum

### 3.2. Electron beams: possible equivalency of exited electrons states in a beam and in β-decay.

There is certain analytic interest in β-decay processes with asymmetry of emitted electrons [10], due to (as it is supposed to be) parity violation in weak interactions. β - asymmetry in angular distribution of electrons was registered for the first time during experiments with polarized nucleuses $_{27}Co^{60}$, β-spectrum of which is characterized by energies of MeV. If in the process of β-decay exited electrons are born, then along with decay scheme

$$n \rightarrow p + e^- + \tilde{v} \tag{11}$$

there will be also decay scheme

$$n \rightarrow p + (e^*)^- + \tilde{v} \rightarrow e^- + \tilde{\gamma} + \tilde{v} \tag{12}$$

where $\tilde{\gamma}$ is graviton.

Decay (12) is energetically limited by energy values of 1 MeV order (in rough approximation), taking into consideration that the difference between lower excitation level of electron's energy (in own gravitational field) and general <100 keV and the very character β-spectrum. Consequently, $_{27}Co^{60}$ nucleuses decay can proceed with equal probability as it is described in scheme (11) or in scheme (12). For the light nucleuses, such as $_1H^3$ β-decay can only proceed as it is described in scheme (11). At the same time, emission of graviton by electron in magnetic field can be exactly the reason for β-asymmetry in angular distribution of electrons. If so, then the phenomenon of β-asymmetry will not be observed in light β-radioactive nucleuses. This would mean that β-asymmetry in angular distribution of electrons, which is interpreted as parity violation, is the result of electron's gravitational emission, which should be manifested in existence of lower border β-decay, as that's where β-asymmetry appears to be. On the other hand, if in process of β-decay creation of electrons' exited states actually takes place, there further emission leads asymmetry in angular distribution of electrons, then it can be proven in simpler experiments with electron beams, using anode diaphragm to achieve exited states. Thereby two aims can be reached simultaneously:

a) Receiving additional information about the character of stationary states in own gravitational field spectrum.

b) Registering the angular distribution of electrons character (magnetic field oriented) with presence of exited states and without them will answer the question of electrons' exited states role in possible asymmetry of angular distribution.

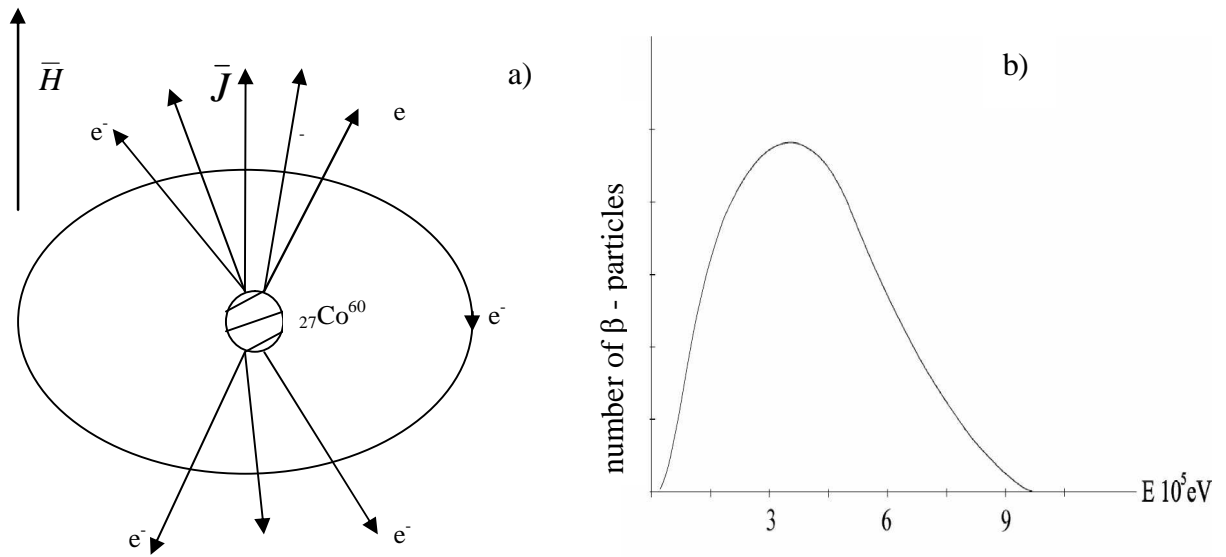

Figure 3. a) Scheme of experiment to identify parity violation in β-decay, b) an example of β spectrum, obtained during experiments; $\bar{H}$ - circular current magnetic field, $\bar{J}$ - cobalt isotope nuclear spin $_{27}Co^{60}$.

Experimental measurements of electrons' angular distribution (after passing anode diaphragm) in the field of electrons' exited states (with accelerating potential values ~ 40 kV and lower) are standard and are necessary for comparative analyses. On Fig. 3 as a visual aid there are a) scheme of experiments By, b) example of β spectrum. With the presence of electrons' exited states (in own gravitational field) to situations can take place:

1) Electrons in exited states can be registered on detector.

2) Exited states decay in area close to anode diaphragm (in case of β-decay in β type emitting area).

In respect of β-decay, comparison of decay data (figures 3a, 3b) gives ground for the question that presence of electrons' exited states (created in nucleus at the moment of β – decay) with certain angular quantum value can be the reason for asymmetry in angular distribution of electrons, or it can be recoil after electrons' exited states decay: curve 3b is the evidence of presence of a certain number of electrons, for which recoil energy during the decay (6 – 30) keV is substantive.

**Conclusion.**

1. Identification of electrons' gravitational emission spectrum lines can be carried out during experiments with electron beams on spectrum lines that are analogous to the lines of characteristic electromagnetic emission of target in wave length range (0,01 – 0,04)Å.

**2.** Electrons' states in β-decay can be analogous to exited electrons' states in own gravitational field; it can be verified by registration of electrons distribution over energy range after electron beam passes foil anode diaphragm with values of accelerating potential of the beam, necessary to provide exited states if the electrons in energy range ~(40 – 171) keV and without them (with lower values of accelerating potential).

**R e f e r e n c e s**

1. Hilbert D. Grundlagen der Physik, 1 Mitt; Gott. Nachr.,1915, math.-nat.kl., 395.
2. Fisenko S. et al., Phys. Lett. A, 148,8,9 (1990) 405.
3. Fisenko S., Fisenko I., PCT Gazette № 46(2005) 553 (IPN WO2005/109970A1).
4. Fisenko S., Fisenko I. http://www.arxiv.org/abs/physics/0604047 (2006)
5. Siravam C. and Sinha K., Phys. Rep. 51 (1979) 112.
6. Friedmann A. Zs.Phys. 10, 377 (1922)
7. W. Pauli: Theory of Relativity; Pergamon Press; 1958.
8. M. Г. Haines, P. D. LePell, C. A. Coverdale, B. Jones, C. Deeney, J. P. Apruzese Phys. Rev. Lett., 96, 075003 (2006)
9. Fisenko S. http://arxiv.org/abs/gr-qc/0105035 (2001)
10. Shpolsky E.V. Atomic physics; v. 1, Nauka, Moskow, 1974.
11. Wu Z. S., Moshkovsky S. A., β-Decay, Atomizdat, Moscow (1970).